\documentclass[conference]{IEEEtran}
\IEEEoverridecommandlockouts
\usepackage{cite}
\usepackage{subcaption}
\usepackage{xspace}
\usepackage{graphicx}
\DeclareCaptionLabelSeparator{custom}{. }
\usepackage{amsmath,amssymb,amsfonts}
\usepackage{algorithmic}
\usepackage[moderate,tracking=normal]{savetrees}
\usepackage{graphicx}
\usepackage{gensymb}
\usepackage{url}
\usepackage{textcomp}
\usepackage{xcolor}
\def\BibTeX{{\rm B\kern-.05em{\sc i\kern-.025em b}\kern-.08em
    T\kern-.1667em\lower.7ex\hbox{E}\kern-.125emX}}
    
\usepackage{mathrsfs}
\usepackage{amssymb}% http://ctan.org/pkg/amssymb
\usepackage{pifont}% http://ctan.org/pkg/pifont
\usepackage{multirow}
\usepackage{blindtext}
\usepackage{graphicx} % if you want to include jpeg or pdf pictures

\usepackage{mathtools}
\usepackage{lipsum, color}

\usepackage{amssymb}
\usepackage{amsthm}
\usepackage{upgreek}
\usepackage{algorithm}
\usepackage{algorithmic}
\usepackage{bbold}
\usepackage{dsfont}
\usepackage{cleveref}
\usepackage[english]{babel}
\usepackage[autostyle]{csquotes}
\usepackage{bigints}
\graphicspath{ {images/} }
\usepackage{float}
\usepackage{titlesec}
\titleclass{\subsubsubsection}{straight}[\subsection]
\usepackage[numbers,sort&compress]{natbib}
\usepackage[labelfont=bf]{caption}
\usepackage{caption}
\usepackage[justification=centering]{caption}
\usepackage[justification=centering]{subcaption}
\usepackage{subcaption}
\usepackage{cool}
\usepackage{booktabs}
\usepackage[english]{babel}

\usepackage{authblk}
\author[1]{\normalsize Dhiraj Bhattacharjee}
\author[1]{Aizaz U. Chaudhry}
\author[1]{Halim Yanikomeroglu}
\author[2]{Peng Hu}
\author[3]{Guillaume Lamontagne\vspace{-1em}}

\affil[1]{Department of Systems and Computer Engineering, Carleton University, Ottawa, ON K1S 5B6, Canada}
\affil[2]{National Research Council Canada (NRC), 1200 Montreal Road, Ottawa, 
ON K1A 0R6, Canada}
\affil[3]{MDA, Sainte-Anne-de-Bellevue, QC H9X 3R2, Canada}
{
    \makeatletter
    \renewcommand\AB@affilsepx{\protect\Affilfont}
    \makeatother

    \affil[ ]

    \makeatletter
    \renewcommand\AB@affilsepx{, \protect\Affilfont}
    \makeatother

    \affil[1]{\{dhirajbhattacharjee, auhchaud, halim\}@sce.carleton.ca}
    \affil[2]{peng.hu@nrc-cnrc.gc.ca}
    \affil[3]{guillaume.lamontagne@mda.space\vspace{-1.6em}}
}

\begin{document}

\title{Laser Inter-Satellite Link Setup~Delay: Quantification, Impact, and Tolerable Value\vspace{-0.4em}\\
% {\footnotesize \textsuperscript{*}Note: Sub-titles are not captured in Xplore and
% should not be used}
% \thanks{Identify applicable funding agency here. If none, delete this.}
}

% \author{\IEEEauthorblockN{Dhiraj Bhattacharjee, Aizaz U. Chaudhry, and Halim Yanikomeroglu}
% \IEEEauthorblockA{\textit{Department of Systems and Computer Engineering, Carleton University, Ottawa, ON K1S 5B6, Canada} \\
% % \textit{Carleton University}\\
% % Ottawa, ON K1S 5B6, Canada \\
% \{dhirajbhattacharjee, auhchaud, halim\}@sce.carleton.ca}
% % \and
% % \IEEEauthorblockN{Aizaz U. Chaudhry}
% % \IEEEauthorblockA{\textit{Department of Systems and Computer Engineering} \\
% % \textit{Carleton University}\\
% % Ottawa, Canada \\
% % auhchaud@sce.carleton.ca}
% % \and
% % \IEEEauthorblockN{Halim Yanikomeroglu}
% % \IEEEauthorblockA{\textit{Department of Systems and Computer Engineering} \\
% % \textit{Carleton University}\\
% % Ottawa, Canada \\
% % halim@sce.carleton.ca}

% \vspace{-2 em}}

\maketitle

\begin{abstract}
Dynamic laser inter-satellite links (LISLs) provide the flexibility of connecting a pair of satellites as required (dynamically) while static LISLs need to be active continuously between the energy-constrained satellites. However, due to the LISL establishment time (termed herein as LISL setup delay) being in the order of seconds, realizing dynamic LISLs is currently unfeasible. Towards the 
realization of dynamic LISLs, we first study the quantification of LISL setup delay; then we calculate the end-to-end latency of a free-space optical satellite network (FSOSN) with the LISL setup delay; subsequently, we analyze the impact of LISL setup delay on the end-to-end latency of the FSOSN. We also provide design guidelines for the laser communication terminal manufacturers in the form of maximum tolerable value of LISL setup delay for which the FSOSN based on Starlink's Phase I satellite constellation will be meaningful to use for low-latency long-distance inter-continental data communications.
\end{abstract}
\vspace{-0.6em}
\section{Introduction\vspace{-0.6em}}
In recent advancements of wireless communication of 6G era, satellite networks have been seen as an integral part along with terrestrial networks for global broadband coverage specially for enabling broadband Internet in rural and remote areas \cite{digitaldivide}, low-latency long-distance inter-continental data communications \cite{chaudhry2021optical}, and IoT based monitoring and remote surveillance \cite{energydhiraj}. From 3GPP definition, satellite payloads could either be transparent or regenerative \cite{3gpp}. In transparent scenario, inter-continental communication has to go up (ground station to satellite) and down (satellite to ground station) frequently to reach from source to destination. With regenerative payload, communication between satellites over inter-satellite links (ISLs) could be a better option in such long-distance communication. Compared to RF-based ISLs, laser ISLs (LISLs) have the advantage of higher bandwidth, smaller antenna size, higher directivity, less power consumption, less chance of interception and interference, etc \cite{freespaceoptics}. Exploiting these LISLs in low Earth orbit (LEO) or very low Earth orbit (VLEO) satellite mega constellations, free-space optical satellite networks (FSOSNs) can be realized in space \cite{9393372}.

On the basis of an LISL's active duration, LISLs can be classified into two types: static LISLs and dynamic LISLs. Static LISLs are those LISLs which are kept active all the time, e.g., SpaceX's Starlink will have four static LISLs per satellite which will be operating all the time \cite{spacex}. In contrast, dynamic LISLs can be established dynamically between satellites (which are within the LISL range) at any time on demand depending upon data communication requirements. To realize such dynamic LISLs instantaneously, we need to have very precise and efficient pointing, acquisition, and tracking (PAT) system \cite{kaushal2017free}. 

Before two satellites start communicating via LISLs, transmitting satellite needs to position its laser beam within the field of view of receiver satellite (pointing). Then the receiver satellite needs to align itself towards the arriving beam (acquisition). Finally, transmitter and receiver continue this process as the communication goes on (tracking) \cite{PAT}. Now, we define LISL setup delay as the time taken by the PAT system to establish the LISL, i.e., the sum of pointing time and acquisition time. This delay will be introduced to the end-to-end latency from a source ground station to destination ground station when the path over an FSOSN changes. Note that when the path changes, it could lead to one or multiple new LISLs. However, LISL setup delay will be introduced only once as multiple LISLs can be established simultaneously during a time slot.

Satellites are driven by onboard battery and solar power, and satellite battery power is a very precious resource, which should be used intelligently. On that regard, static LISLs are always active whether they are being used or not. This will drain the satellite battery and satellites could be dead more often and they need to be de-orbited and new satellites have to be launched. This in turn will increase the maintenance expenditure. On the other hand, dynamic LISLs will be an energy efficient approach where LISLs are only established as required. With dynamic LISLs, two neighbour satellites could connect whenever they are within LISL range and this will provide more routing options. These links between neighbour satellites could be inter-orbital plane, crossing orbital plane, inter-shell, and even inter-constellation (e.g., between Starlink and OneWeb). Also, as the LEO/VLEO satellites are mobile, communications between satellites and ground stations will always be through dynamic laser links. Furthermore, in an operating satellite constellation, if one or many satellites fail, dynamic LISLs will instantaneously reroute the traffic by avoiding the dead satellite(s).

LISL setup delay for current laser communication terminals (LCTs) varies from few seconds to tens of seconds \cite{carrizo2020optical}. This prevents us from realizing dynamic LISLs in next-generation FSOSNs (NG-FSOSNs) in late 2020s. In next-next-generation FSOSNs (NNG-FSOSNs) (in 2030s), due to advancement in satellite PAT technology, LISL setup delay could be reduced to the order of a few milliseconds and dynamic LISLs could become a reality. In this context, we study the quantification of LISL setup delay in the FSOSN based on Starlink's Phase I constellation \cite{spacex}. We calculate the end-to-end latency of this FSOSN using different values of the LISL setup delay in different inter-continental connection scenarios and different LISL ranges for satellites. We investigate the impact of LISL setup delay on overall latency and provide design guidelines for LCT manufacturers to leverage full potential of NNG-FSOSNs via dynamic LISLs. To the best of our knowledge, there exists no study on LISL setup delay that examines its quantification, and its impact on end-to-end latency along with its maximum tolerable values.

The authors of \cite{delaynotoptional} state that for terrestrial distances larger than 3000 km, FSOSNs could provide a better latency performance as compared to the optical fiber terrestrial network (OFTN). In high-frequency trading of stocks, even 1 ms improvement in latency could generate \$100 million of revenue per year \cite{whentocrossover}. Thus, in such long distance-communication, FSOSNs could be a better solution compared to the OFTN. With this objective, we come up with maximum tolerable values of LISL setup delay for which latency performance of the FSOSN based on Starlink's Phase I constellation will be better than the OFTN. This maximum value of LISL setup delay can be a design guideline for LCT manufacturers to leverage advantages of dynamic LISLs in NNG-FSOSNs.

The paper organization is as follows. We discuss related work on network latency of satellite networks and examine LISL setup delays of current LCT manufacturers in Section \ref{rw}. In Section \ref{method}, we elaborate on how we quantify LISL setup delay, calculate end-to-end latency, and define performance metrics. We present our results in Section \ref{res}, discuss insights and design guidelines in Section \ref{insight}, and conclude our discussion with some possible future extensions in Section \ref{con}.\vspace{-0.65em}
\section{Related Work}\label{rw}
Currently, Mynaric's LCT CONDOR needs 30 seconds to establish an LISL between two satellites for the first time. Once the orbital parameters are exchanged between satellites, it takes 2 seconds to setup an LISL for every next time \cite{carrizo2020optical}. Tesat \cite{tesat} and General Atomics \cite{ga} have LCTs for LEO/VLEO constellations which have LISL setup delay in the range of tens of seconds.

In any communications network, the end-to-end latency from source to destination typically has four components: propagation delay, transmission delay, queuing delay, and processing delay \cite{kurose2010computer}. Based on this latency model, authors of \cite{chaudhry2021optical} compared the latency performance of FSOSNs and OFTNs. As stated earlier that latency-wise, FSOSNs can be a better alternative to OFTN for longer communication distances \cite{delaynotoptional}. On that regard, authors of \cite{whentocrossover} and \cite{chaudhry2022crossover} have come up with a concept of crossover distance to determine that for a certain terrestrial distance, which one will provide a better latency performance among FSOSN and OFTN. 

Authors of \cite{10.1145/3365609.3365859} have suggested ground stations as relays as a substitute of ISLs where satellites have transparent payload. With this network architecture, they proved that constellations like Starlink can provide better latency performance compared to OFTN. In addition to that, idle user terminals can also be used which will provide further improvement in latency performance. However, \cite{withoutisl} shows that exploiting ISLs can reduce variation in latency performance as well as reduce the effects of weather impairments. To analyze network delay, authors of \cite{9149009} have modeled each satellite node as M/M/1 queue in a multihop scenario where each satellite can receive packets from ground station as well as other satellite node. Authors of \cite{chaudhry2022temporary} highlighted the importance of temporary LISLs (defined as LISLs which are established temporarily with satellites that are within LISL range) in order to achieve better latency performance compared to static LISLs in FSOSNs. They showed that with temporary LISLs, there exist more number of LISLs which will provide better routing options. They also reported that temporary LISLs are more useful at lower LISL ranges. Authors of \cite{whentocrossover} and \cite{chaudhry2022temporary} mentioned about LISL setup delay in FSOSNs. With that respect, in FSOSNs, along with the other four end-to-end latency components discussed earlier, we introduce LISL setup delay to the latency model. 
\begin{table*}[ht]
\caption{$\eta_{LE}(without\;\; \eta_s)$,~$\alpha_i$,~$\eta_s$,~and~$\eta_{LE}(with\;\; \eta_s)$~of~the~shortest~paths~at~first~6~time~slots ~over~the~FSOSN~for~New~York--Istanbul~inter-continental~connection.}
\centering
\label{strands}
\begin{tabular}{|c| c| c| c| c| c|}
\hline
 &  & $\eta_{LE}$&  &  & $\eta_{LE}$\\
Time & Shortest &(without & $\alpha_i$ & $\eta_s$ &(with\\
Slot & Path & $\eta_s$) & & (ms) &$\eta_s$) \\
 & & (ms) & & & (ms)\\
\hline
1 & GS at New York, satellite \emph{x10919, x11115, x11312, x11509, x11609, x11611, x12166}, GS at Istanbul & 38.09 & 0 & 0 & 38.09\\
\hline
2 & GS at New York, satellite \emph{x11503, x11505, x11507, x11509, x11609, x11611, x12166}, GS at Istanbul & 38.08 & 1 & 100 & 138.08\\
\hline
3 & GS at New York, satellite \emph{x11503,x11505, x11507, x11509, x11609, x11611, x12166}, GS at Istanbul & 38.07 & 0 & 0 & 38.07\\
\hline
4 & GS at New York, satellite \emph{x11503,x11505, x11507, x11509, x11609, x11611, x12166}, GS at Istanbul & 38.06 & 0 & 0 & 38.06\\
\hline
5 & GS at New York, satellite \emph{x11503,x11505, x11507, x11509, x11609, x11611, x12166}, GS at Istanbul & 38.05 & 0 & 0 & 38.05\\
\hline
6 & GS at New York, satellite \emph{x11503, x11505, x11507, x11508, x11608, x11903, x12166}, GS at Istanbul & 38.00 & 1 & 100 & 138.00\\
\hline
% 7 & GS at New York, satellite \emph{x11503, x11505, x11506, x11508, x11608, x11903, x12166}, GS at Istanbul & 38.00 & 1 & 100 & 138.00\\
% \hline
% 8 & GS at New York, satellite \emph{x11503, x11504, x11506, x11508, x11608, x11903, x12166}, GS at Istanbul & 37.99 & 1 & 100 & 137.99\\
% \hline
% 9 & GS at New York, satellite \emph{x11503, x11504, x11506, x11508, x11608, x11903, x12166}, GS at Istanbul & 37.98 & 0 & 0 & 37.98\\
% \hline
% 10 & GS at New York, satellite \emph{x11503, x11504, x11506, x11508, x11608, x11903, x12166}, GS at Istanbul & 37.98 & 0 & 0 & 37.98\\
% \hline
\end{tabular}
\end{table*}
\section{Methodology}\label{method}
To quantify LISL setup delay, calculate end-to-end latency from source to destination with the LISL setup delay, investigate the impact of LISL setup delay on overall latency, and present design guidelines for LCT as a form of maximum tolerable value of LISL setup delay, we simulate Starlink's Phase I Version 2 constellation in AGI's Systems Tool Kit (STK) platform \cite{stk}. This constellation has a total of 1584 satellites consisting 24 orbital planes with each of them having 66 satellites \cite{spacex}. The orbits are at an inclination of 53$\degree$ with respect to the equator and satellites are at an altitude of 550 km. With these constellation parameters, we generate this constellation's satellites in STK with a certain LISL range (i.e., the range over which a satellite in an FSOSN can establish an LISL with any other satellite within this range) along with ground stations at New York, London, Istanbul, and Hanoi. Next we extract the data from STK (e.g., vertices, edges, length of edges, etc) at every second for one hour simulation period to Python platform. Then we apply Dijkstra's shortest path algorithm \cite{dijkstra1959note} to find shortest path at every time slot (equal to one second in duration) for the source to destination pairs: New York--London, New York--Istanbul, and New York--Hanoi. 

In our investigation, we consider 4 different values of LISL range: 1500 km, 1700 km, 2500 km, and 5016 km. The minimum range to have communication with nearest neighbor at the immediate left and right orbital planes is 1500 km in Starlink's Phase I Version 2 constellation. At this range, a satellite can connect to two satellites in front and two at rear in the same orbital plane making total 6 connections. At 1700 km range, a satellite can connect to three immediate neighbors on the left, three immediate neighbors on the right, and four intra-orbital plane neighbors making a total of 10 possible connections. The maximum possible LISL range for Starlink's Phase 1 constellation can be calculated as 5016 km \cite{9393372}. The 2500 km LISL range is taken as an intermediate value between 1700 km and 5016 km.  

\subsection{Quantification of LISL Setup Delay}
We define LISL setup delay indicator (a binary variable) as follows: if the shortest paths of $(i-1)^{th}$ time slot and $i^{th}$ time slot are exactly same, no LISL setup delay is to be included in the end-to-end latency, and the LISL setup delay indicator, $\alpha_i$ is $0$. If shortest path changes from $(i-1)^{th}$ to $i^{th}$ time slot, $\alpha_i$ is $1$. Considering $\eta_s$ as LISL setup delay, we denote end-to-end latency without and with LISL setup delay as $\eta_{LE}(without\; \eta_s)$ and $\eta_{LE}(with\; \eta_s)$, respectively. In Table \ref{strands}, we show shortest paths (satellite naming convention follows \cite{9393372}) and corresponding values of $\eta_{LE}(without\; \eta_s)$, $\alpha_i$, and $\eta_{LE}(with\; \eta_s)$ for first 6 time slots over the FSOSN for New York to Istanbul inter-continental connection at an LISL range of 1500 km. From Table \ref{strands}, we can see that shortest path could change from time to time. This is due to the fact that as LEO satellites are moving at high orbital speeds, either a shortest path at one time instance may even not exist in the next time instance (due to one or multiple satellites moving out of range) or there may become available a new shortest path. $\eta_{LE}(without\; \eta_s)$ has two major components: propagation delay and node delay (sum of processing and queuing delay is node delay). We calculate propagation delay as sum of lengths of all the laser links in the shortest path divided by speed of light in vacuum and we consider node delay as 1 ms \cite{1msdelay}. From Table \ref{strands}, it is evident that shortest paths are not same for $1^{st}$ and $2^{nd}$ time slots, so $\alpha_{2}=1$. Considering $\eta_s=100$ ms, $\eta_{LE}(with\; \eta_s)$ will be 38.08+100, i.e., 138.08 ms at time slot 2. The shortest path remains unchanged from time slot 3 to 5, i.e., $\alpha_{3}=\alpha_{4}=\alpha_{5}=0$ and corresponding $\eta_{LE}(with\; \eta_s)$ values remain same as $\eta_{LE}(without\; \eta_s)$. Then again at time slot 6, shortest path changes which makes $\alpha_{6}=1$.

\subsection{Path Change Rate}We simulate for 3600 time slots, one time slot being equal to 1 second in duration and we define the path change rate, $\lambda$ as the average number of instances the shortest path from source to destination changes (represented in percentage) and mathematically it can be calculated as
\begin{equation}\label{path_change_rate}
    \lambda=\frac{1}{3600} \sum_{i=1}^{3600} \alpha_i\;\;\times100\%.
\end{equation}

% \begin{equation}\label{path_stability_rate}
%     \mu=1-\lambda
% \end{equation}

\subsection{End-to-End Latency} Averaging $\eta_{LE}(with\; \eta_s)$ and $\eta_{LE}(without\; \eta_s)$ over 3600 time slots, we get average end-to-end latency with and without $\eta_s$ as $\overline{\eta_{LE}}(with\; \eta_s)$ and $\overline{\eta_{LE}}(without\; \eta_s)$, respectively. They are related to $\lambda$ and $\eta_s$ as follows:
\begin{equation}\label{end_to_end_delay}
    \overline{\eta_{LE}}(with\; \eta_s)=\overline{\eta_{LE}}(without\; \eta_s)+\frac{\lambda}{100}\;\eta_s.
\end{equation}

\subsection{Impact of $\eta_s$}
To measure the impact of LISL setup delay, $\eta_s$ on average end-to-end latency, we define $\beta$ as percentage of delay introduced due to $\eta_s$ in average end-to-end latency and calculate it as follows:
\begin{equation}\label{impact}
    \beta=\frac{\overline{\eta_{LE}}(with\; \eta_s)-\overline{\eta_{LE}}(without\; \eta_s)}{\overline{\eta_{LE}}(with\; \eta_s)} \times 100\%.
\end{equation}
\subsection{Tolerable Value of $\eta_s$}
For an inter-continental connection, it is meaningful to use the FSOSN only when $\overline{\eta_{LE}}(with\; \eta_s)$ is lesser than end-to-end latency of the OFTN, $\eta_{LE,\;OFTN}$. Using (\ref{end_to_end_delay}), the following can be written,
\begin{equation}\label{tolerable_1}
    \overline{\eta_{LE}}(without\; \eta_s)+\frac{\lambda}{100}\;\eta_s \leq \eta_{LE,\;OFTN}.
\end{equation}
Now we define the maximum tolerable value of LISL setup delay, $\eta_{s,\textrm{max}}$ as the maximum value of $\eta_s$ so that the average end-to-end latency of the FSOSN is lesser or equal to that of the OFTN and calculate it from (\ref{tolerable_1}) as follows:
\begin{equation}\label{tolerableeqn}
    \eta_{s,\textrm{max}}=\frac{\eta_{LE,\;OFTN}-\overline{\eta_{LE}}(without\; \eta_s)}{\lambda/100}.
\end{equation}
To calculate $\eta_{LE,\;OFTN}$, first we determine the distance from the source ground station to the destination ground station along the surface of the Earth using Haversine formula \cite{WinNT} from latitudes and longitudes of source and destination ground stations. Later we find $\eta_{LE,\;OFTN}$ as that distance divided by speed of light in the optical fiber (having refractive index=1.4675), i.e., 204,287,876 m/s.\vspace{-0.4em}
\section{Results}\label{res}
\begin{figure*}
    \centering
    \begin{subfigure}[b]{0.326\textwidth}
        \centering
        \includegraphics[width=\textwidth]{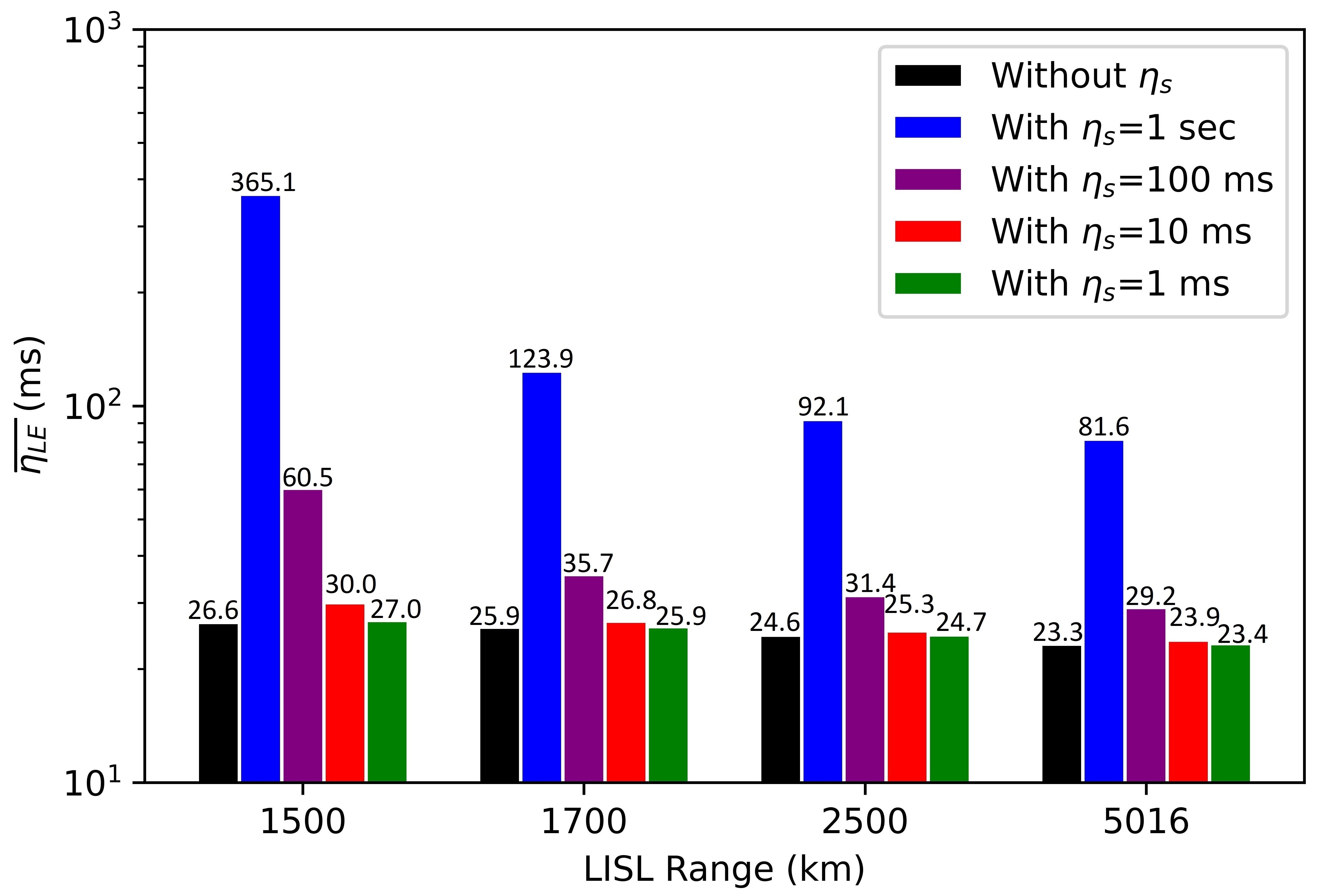}
        \caption{New York to London.}
        \label{latencylondon}
    \end{subfigure}
    % \quad
    \begin{subfigure}[b]{0.326\textwidth}
        \centering 
        \includegraphics[width=\textwidth]{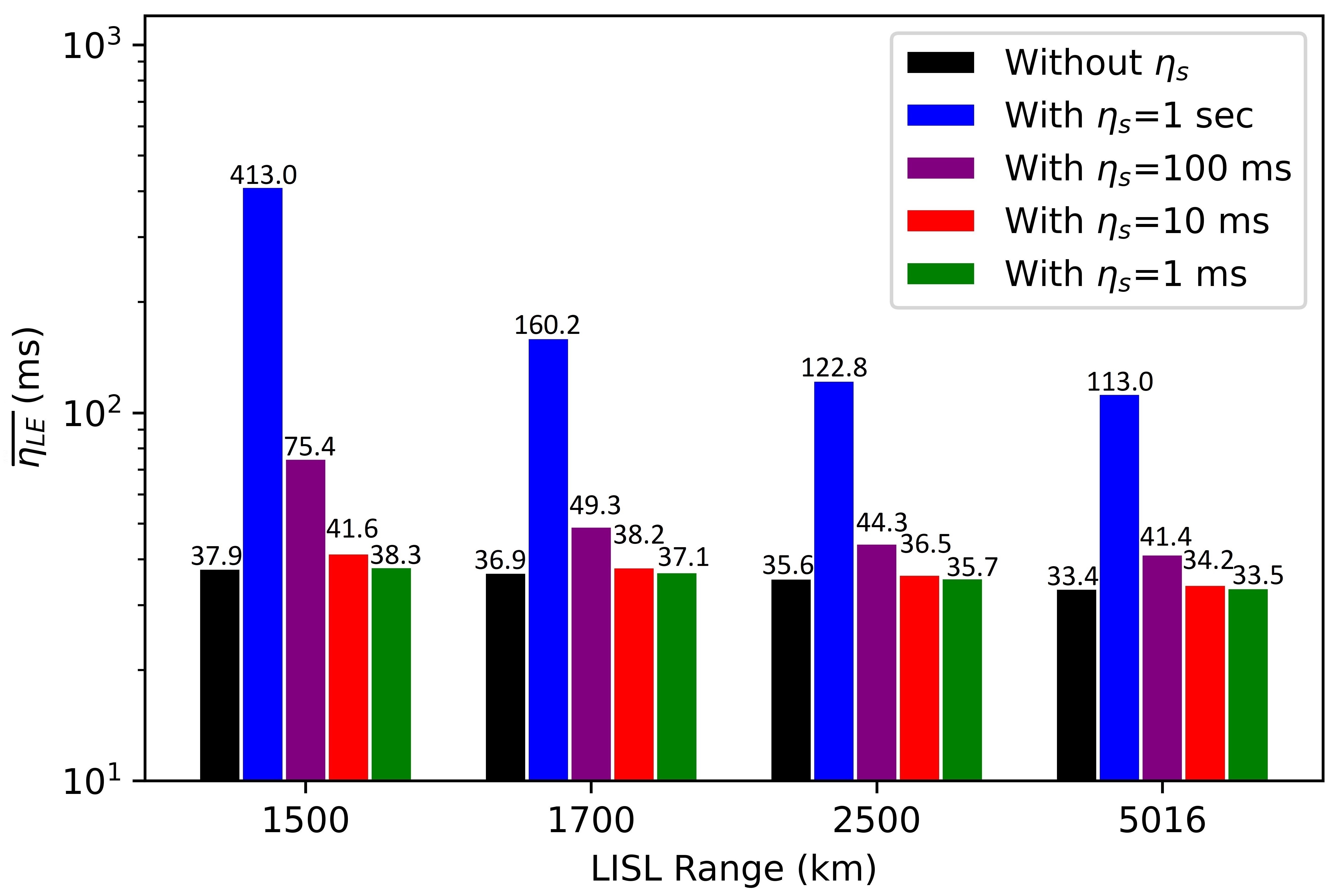}
        \caption{New York to Istanbul.}
        \label{latencyist}
    \end{subfigure}
    % \quad
    \begin{subfigure}[b]{0.33\textwidth}
        \centering 
        \includegraphics[width=\textwidth]{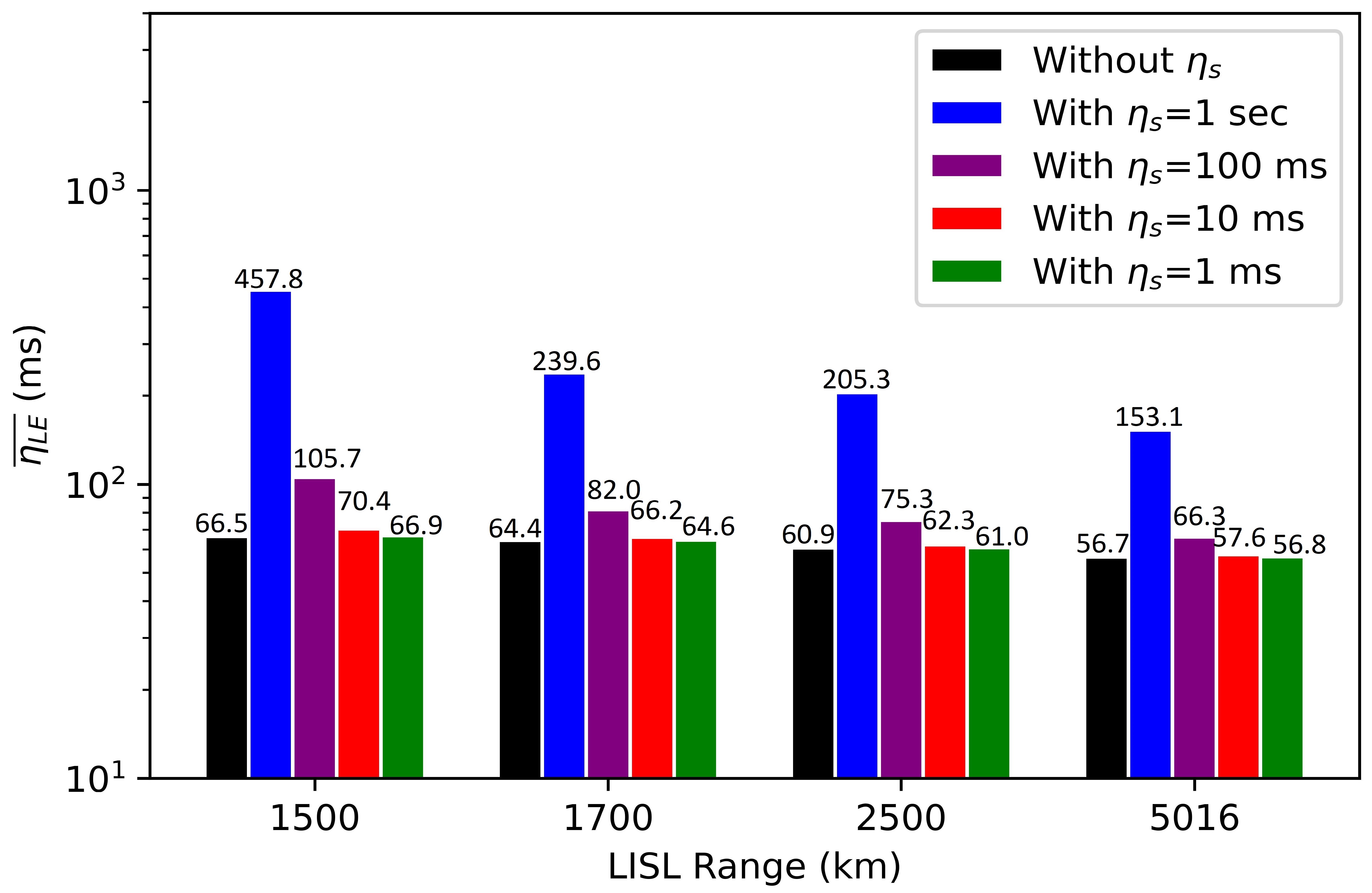}
        \caption{New York to Hanoi.}
        \label{latencyhanoi}
    \end{subfigure}
    % \vskip\baselineskip
    \caption{Average end-to-end latency performance.\vspace{-1.5em}}\label{latency}
\end{figure*}

We consider three inter-continental connections: New York to London (low inter-continental distance connection with terrestrial distance=5593 km), New York to Istanbul (mid inter-continental distance connection with terrestrial distance=8079 km), and New York to Hanoi (high inter-continental distance connection with terrestrial distance=13164 km). For the metrics $\lambda$, $\overline{\eta_{LE}}$, and $\beta$, we show bar plots for the four LISL ranges. To clearly show both high and low values in the same figure, we use log scale in the y-axis in Figs. \ref{pathchange} to \ref{imp}.
\subsection{Path Change Rate}
\begin{figure}
    \centering
    \begin{subfigure}[b]{0.33\textwidth}
    \includegraphics[width=\textwidth]{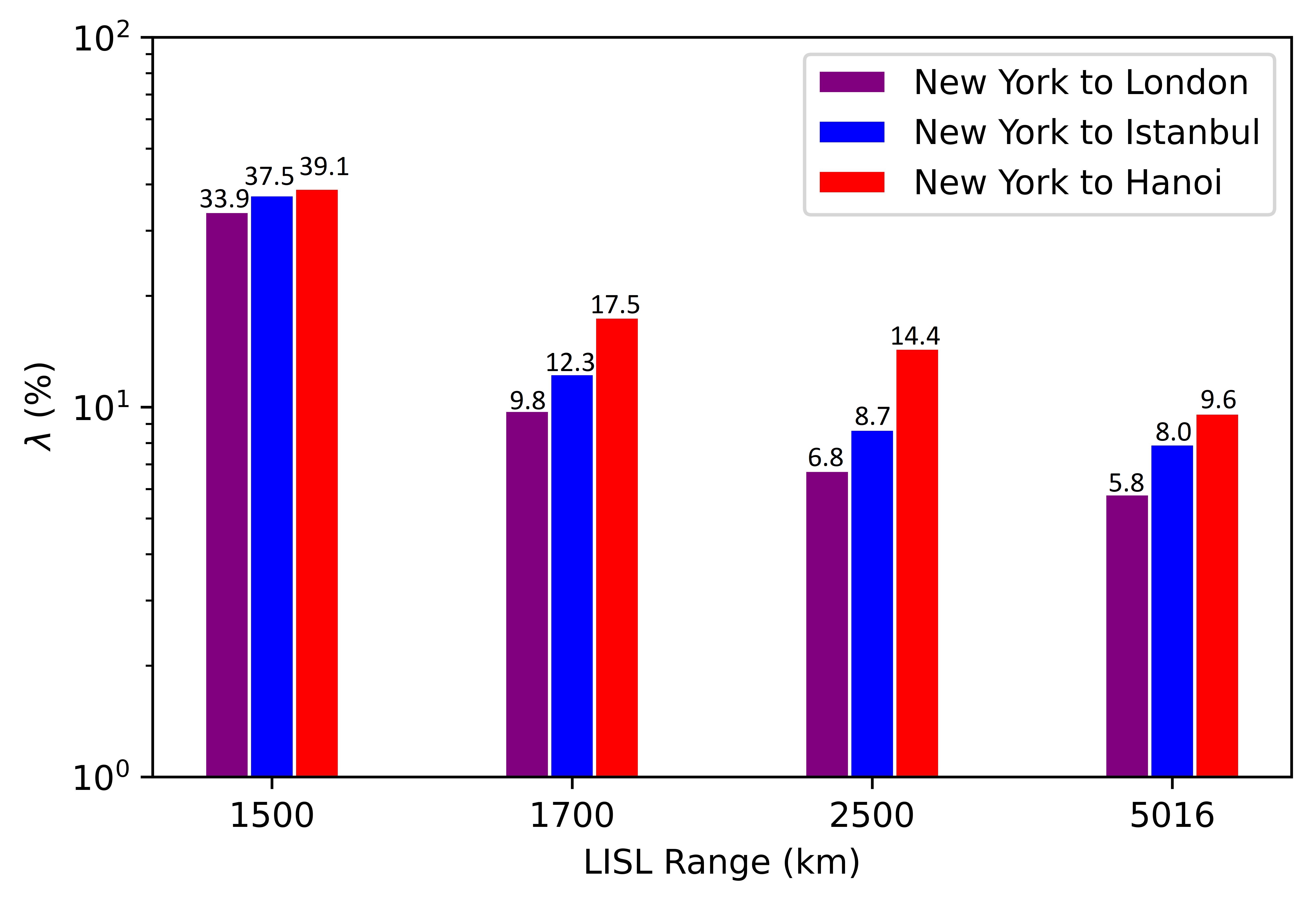}
    \end{subfigure}
    \vspace{-1em}\caption{Path change rate.\vspace{-1.5em}}
    
    \label{pathchange}
\end{figure}

%\subsubsection{Discussion}
In Fig. \ref{pathchange}, we plot $\lambda$ with LISL range varying along x-axis for the three inter-continental connections. For any inter-continental connection, we can observe that $\lambda$ reduces as LISL range increases. Also note that for a particular LISL range, the more the inter-continental distance, the higher the value of $\lambda$. 
\subsection{End-to-End Latency}
Fig. \ref{latency} shows end-to-end latency for the three inter-continental connections averaged over one hour of simulation period without considering $\eta_s$ and with four $\eta_s$ values. As LISL range increases along x-axis, both $\overline{\eta_{LE}} (without\; \eta_s)$ (black bars) and $\overline{\eta_{LE}} (with\; \eta_s)$ (other bars) decrease. For a certain LISL range, the more the value of $\eta_s$, the more the overall latency. For example, in Fig. \ref{latencylondon} with LISL range of 1700 km, $\overline{\eta_{LE}} (with\; \eta_s)$ is 123.9 ms for $\eta_s$=1 sec and it reduces to 35.7 ms when $\eta_s$ is considered to be 100 ms. Also, for a certain LISL range with a certain $\eta_s$ value, the more the inter-continental distance, the more the end-to-end latency for both the cases: $\overline{\eta_{LE}} (without\; \eta_s)$ and $\overline{\eta_{LE}} (with\; \eta_s)$. It is interesting to note that with the increase of LISL range, $\overline{\eta_{LE}} (with\; \eta_s)$ reduces faster compared to $\overline{\eta_{LE}} (without\; \eta_s)$. For example, considering Fig. \ref{latencylondon}, $\overline{\eta_{LE}} (without\; \eta_s)$ drops from 25.9 ms to 24.6 ms when LISL range increases from 1700 km to 2500 km. If we take the ratio and term the ratio as reduction ratio, for this case it will be $\frac{25.9}{24.6}=1.053$. Similarly, for $\overline{\eta_{LE}} (with\; \eta_s)$, the reduction ratio will be $\frac{123.9}{92.1}=1.345$ which is greater than that of $\overline{\eta_{LE}} (without\; \eta_s)$.
\subsection{Impact of $\eta_s$}
\begin{figure*}
    \centering
    \begin{subfigure}[b]{0.31\textwidth}
        \centering
        \includegraphics[width=\textwidth]{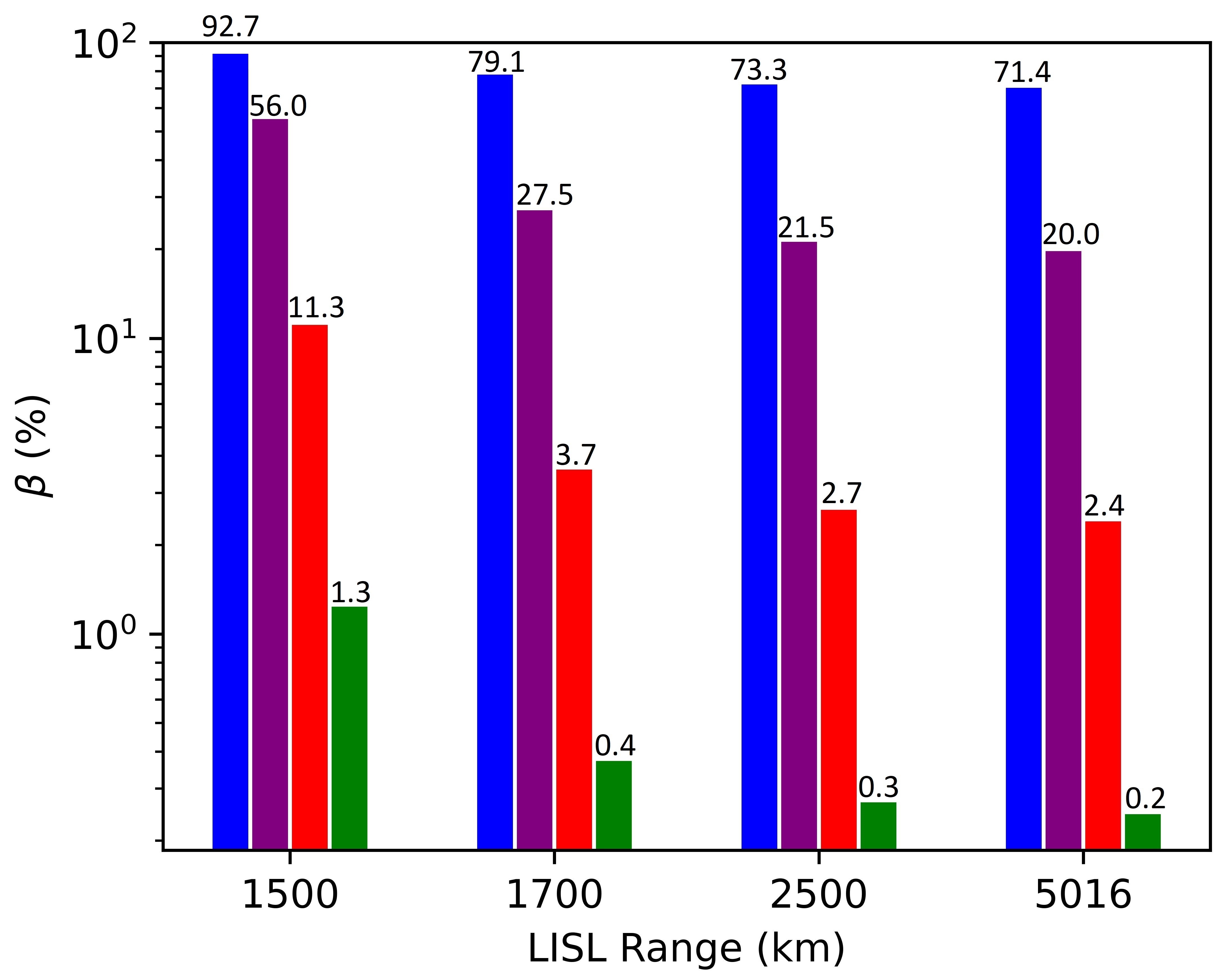}
        \caption{New York to London.}
        \label{impactlondon}
    \end{subfigure}
    % \quad
    \begin{subfigure}[b]{0.365\textwidth}
        \centering 
        \includegraphics[width=5.9cm,height=4.9cm]{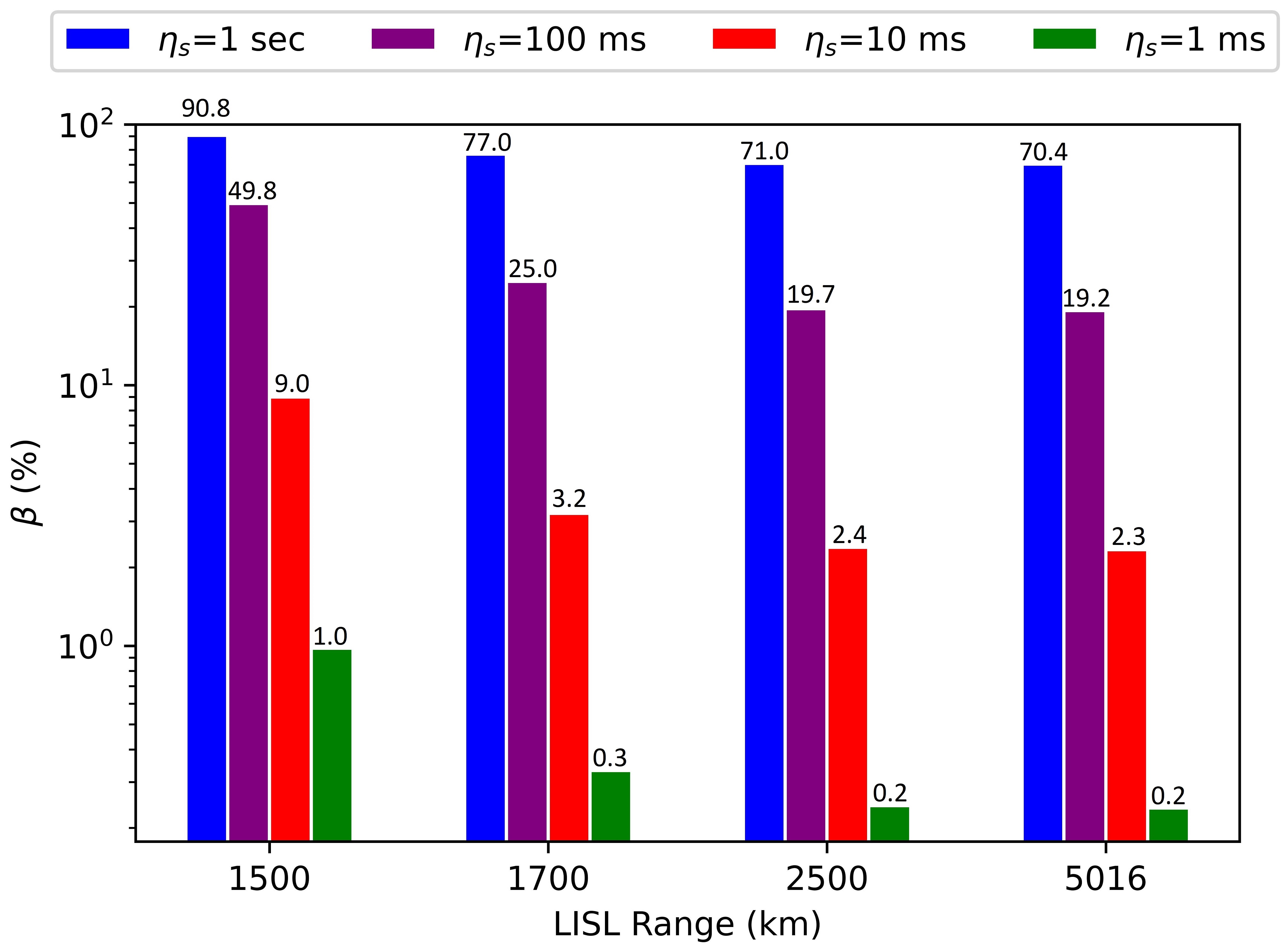}
        \caption{New York to Istanbul.}
        \label{impactist}
    \end{subfigure}
    % \quad
    \begin{subfigure}[b]{0.31\textwidth}
        \centering 
        \includegraphics[width=\textwidth]{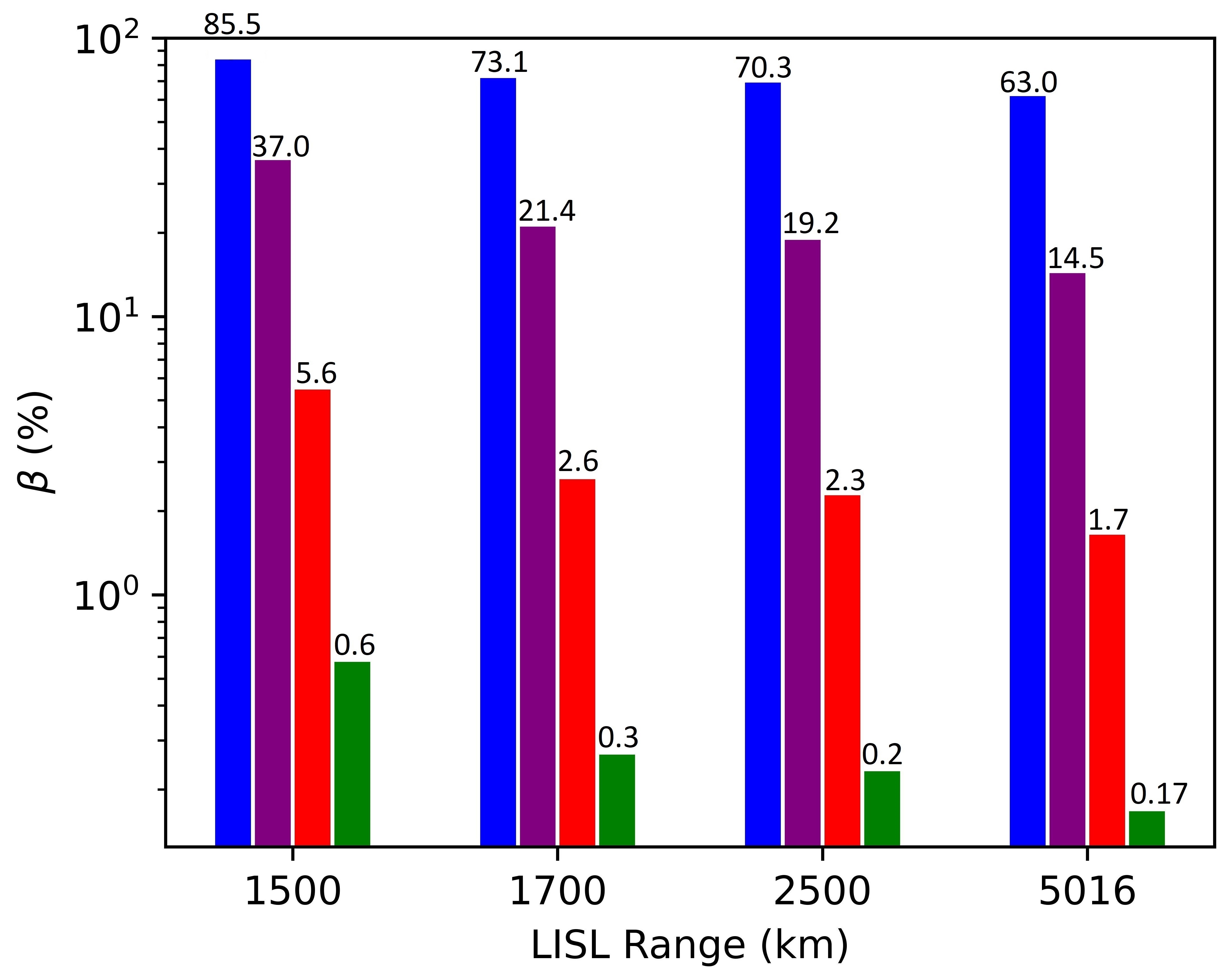}
        \caption{New York to Hanoi.}
        \label{impacthanoi}
    \end{subfigure}
    % \vskip\baselineskip
    \caption{Impact of $\eta_s$ on end-to-end latency.\vspace{-1em}}\label{imp}
\end{figure*}

% \subsubsection{Discussion} 
In Fig. \ref{imp}, we show the variation of $\beta$ with LISL range for four $\eta_s$ values in the three inter-continental connections. As we see, $\beta$ reduces as LISL range increases for a certain $\eta_s$ value. Also, at a certain LISL range, $\beta$ reduces as $\eta_s$ reduces. For example, in Fig. \ref{impactist} with LISL range of 2500 km, $\beta$ is 71\% for $\eta_s$=1 sec. However, when $\eta_s$ reduces to 100 ms, $\beta$ reduces to 19.7\%. In addition, for a certain LISL range with a particular $\eta_s$ value, $\beta$ reduces as inter-continental distance increases. For example, assuming 1700 km of LISL range and $\eta_s$ as 1 sec, $\beta$ reduces from 77\% to 73.1\% when inter-continental connection changes from New York--Istanbul to New York--Hanoi.
\subsection{Tolerable Value of $\eta_s$}
% \subsubsection{Discussion}
In Fig. \ref{tolerablezoom}, we plot $\overline{\eta_{LE}} (with\; \eta_s)$ and $\eta_{LE,OFTN}$ against $\eta_s$. Note that, $\overline{\eta_{LE}} (with\; \eta_s)$ is a straight line with a constant slope and as LISL range increases, the slope reduces. The significance of this figure is where $\overline{\eta_{LE}} (with\; \eta_s)$ for a certain LISL range cuts $\eta_{LE,OFTN}$, the x-coordinate value of the intersection point is $\eta_{s,\textrm{max}}$ as beyond that point, $\overline{\eta_{LE}} (with\; \eta_s)$ will be greater than $\eta_{LE,OFTN}$. To show the intersection points clearly, we only present $\eta_s$ values on the x-axis varying from 1 ms to 100 ms where we mention the coordinates of the intersection points. If we substitute $\eta_{LE,OFTN}=39.55$ ms (from Fig. \ref{tolerablezoomist}), $\overline{\eta_{LE}} (without\; \eta_s)=37.9$ ms (from Fig. \ref{latencyist}), and $\lambda=37.5\%$ (from Fig. \ref{pathchange}) for New York to Istanbul inter-continental connection with 1500 km LISL range in (\ref{tolerableeqn}), we get $\eta_{s,\textrm{max}}$ as 4.4 ms which matches with Fig. \ref{tolerablezoomist}. Also, we should observe from Fig. \ref{tolerablezoom} that, as LISL range increases, $\eta_{s,\textrm{max}}$ also increases. Interesting point to note in Fig. \ref{tolerablezoomhanoi} is that it only shows two intersection points because $\overline{\eta_{LE}} (with\; \eta_s)$ for 1500 km and 1700 km LISL range straight lines (black and blue lines) never intersect with $\eta_{LE,OFTN}$ for $\eta_s>1$ ms values. For 1500 km LISL range, $\overline{\eta_{LE}} (without\; \eta_s)$=66.5 ms (from Fig. \ref{latencyhanoi}) and $\eta_{LE,OFTN}$=64.44 ms (from Fig. \ref{tolerablezoomhanoi}). Putting these values in (\ref{tolerableeqn}), we get negative $\eta_{s,\textrm{max}}$ value which does not exist. Similarly, for 1700 km LISL range, $\overline{\eta_{LE}} (without\; \eta_s)\approx \eta_{LE,OFTN}$ which makes $\eta_{s,\textrm{max}}\approx0$.

\section{Insights and Design Guidelines\vspace{-0.4em}}\label{insight}
\subsection{Insights}\vspace{-0.28em}
\subsubsection{Path Change Rate}
\begin{itemize}
    \item As LISL range increases, there will be lesser hops, i.e., lesser number of satellites for the signal to reach from source to destination. For example, in New York to Istanbul inter-continental connection, average number of hops drops from 7 to 6 when LISL range increases from 1500 km to 1700 km. The lesser the number of hops, the lesser is the chance of a new shortest path. This in turn reduces $\lambda$. Also, when the LISL range increases, two satellites remain in communication range for a longer time span. One of the reasons for the shortest path to change is satellites going out of LISL range, and a shortest path tends to change lesser with longer LISL range. Due to these two reasons, $\lambda$ reduces as LISL range increases.
    \begin{figure*}
    \centering  
     \begin{subfigure}[b]{0.328\textwidth}
        \centering 
        \includegraphics[width=\textwidth]{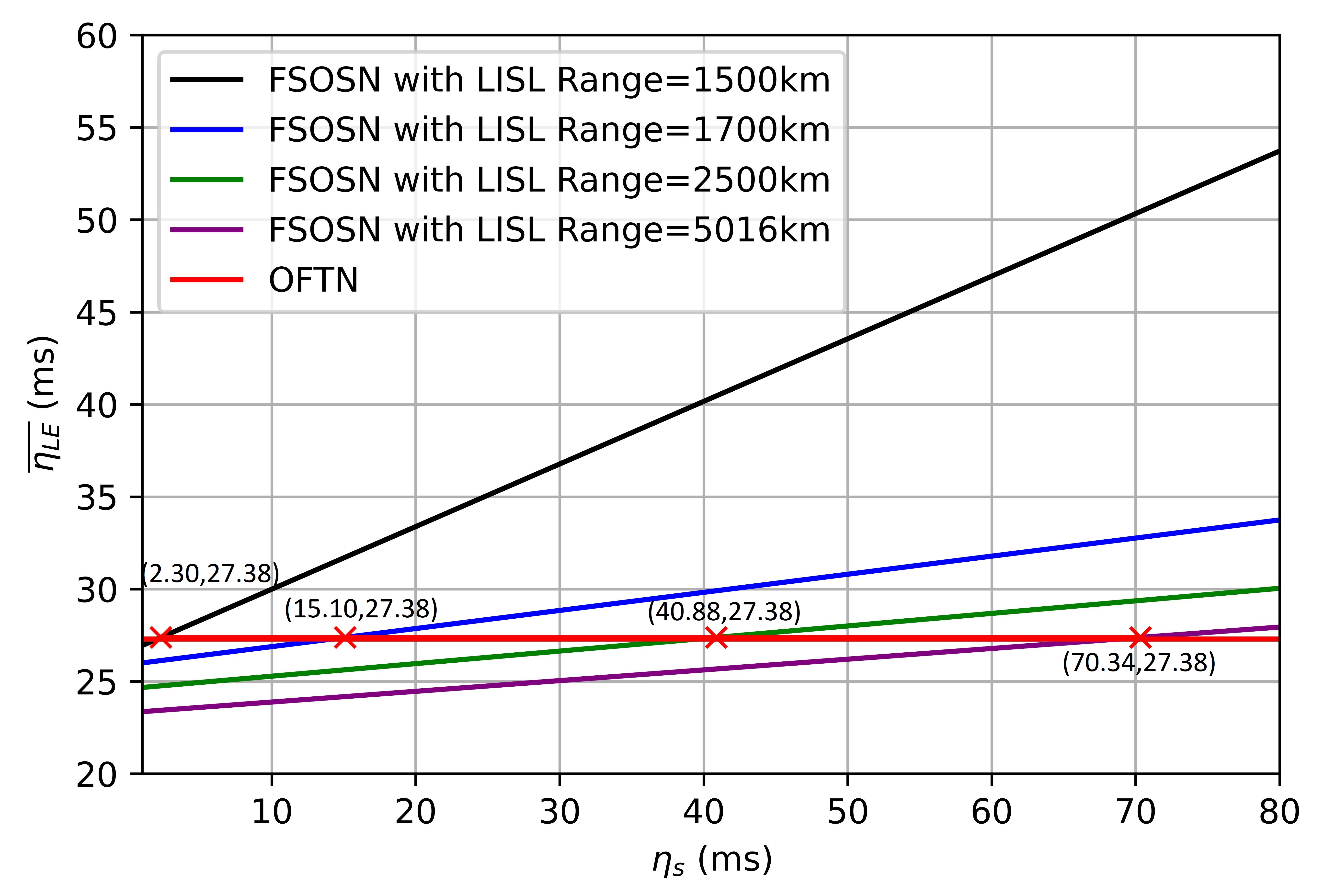}
        \caption{New York to London.}
        \label{tolerablezoomlondon}
    \end{subfigure}
    % \quad
    \begin{subfigure}[b]{0.328\textwidth}
        \centering 
        \includegraphics[width=\textwidth]{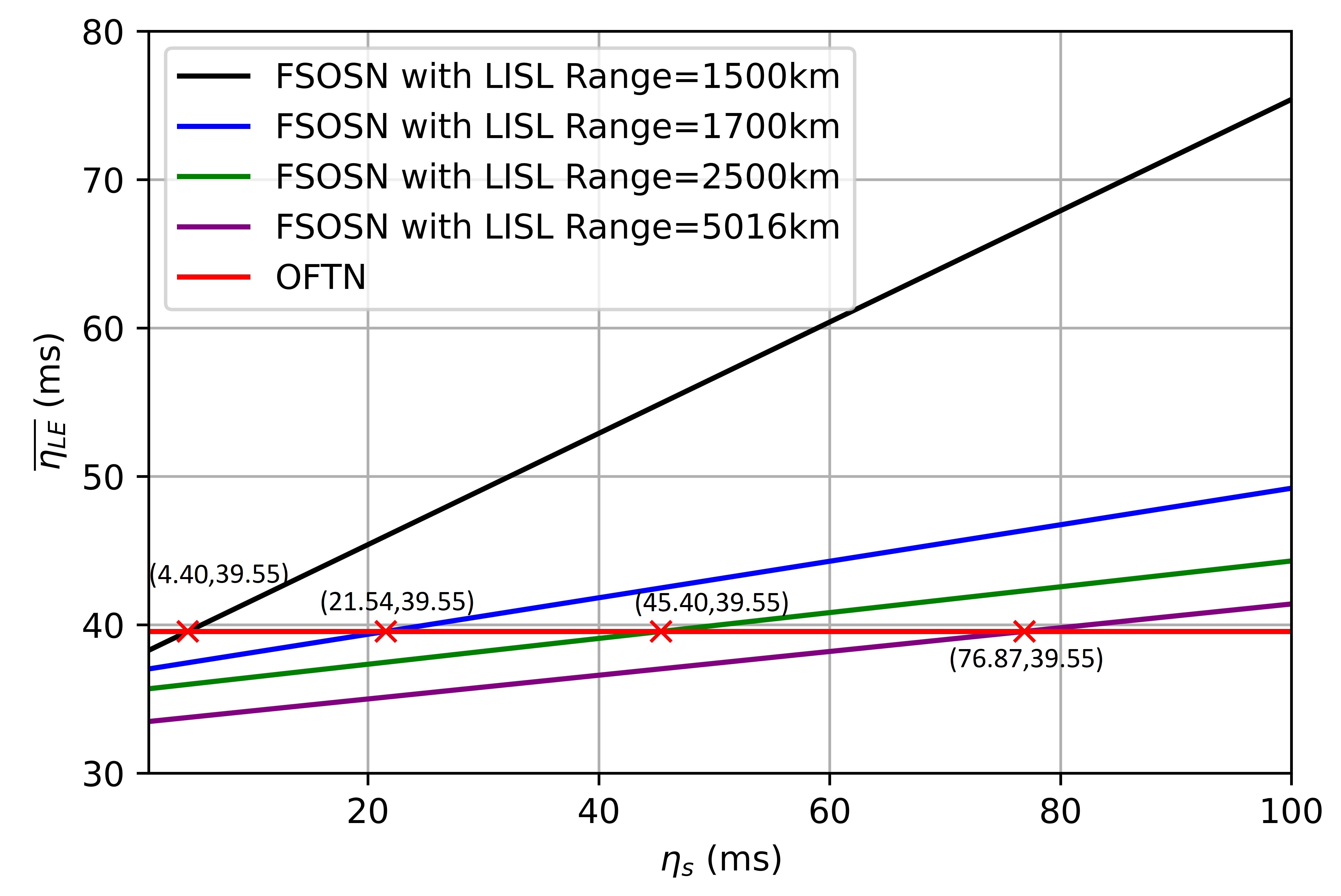}
        \caption{New York to Istanbul.}
        \label{tolerablezoomist}
    \end{subfigure}
    % \quad
    \begin{subfigure}[b]{0.328\textwidth}
        \centering 
        \includegraphics[width=\textwidth]{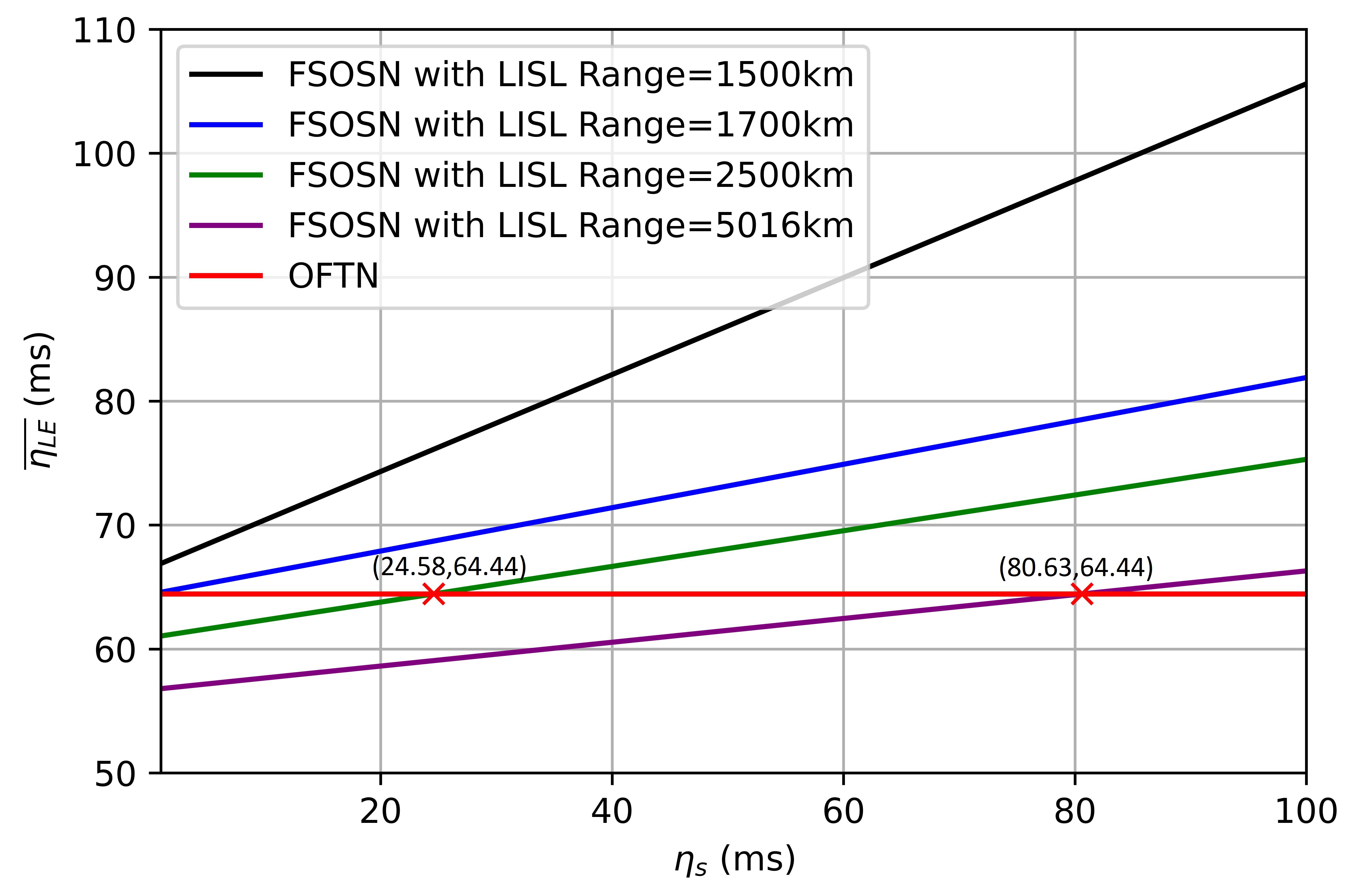}
        \caption{New York to Hanoi.}
        \label{tolerablezoomhanoi}
    \end{subfigure}
    % \vskip\baselineskip
    \caption{Maximum tolerable value of $\eta_s$.}\label{tolerablezoom}
\end{figure*}
    \item For a certain LISL range, the longer the inter-continental distance, the higher the average number of hops which leads to more chances of a new shortest path, and this increases the path change rate, $\lambda$.
    \end{itemize}
\subsubsection{End-to-End Latency}       
  \begin{itemize}
           
    \item Increase in LISL range reduces the number of hops which reduces total node delays. This decreases $\overline{\eta_{LE}} (without\; \eta_s)$ with the increase in LISL range. Now, as both $\lambda$ and $\overline{\eta_{LE}} (without\; \eta_s)$ decrease with the increase of LISL range, from (\ref{end_to_end_delay}), it is clear that $\overline{\eta_{LE}} (with\; \eta_s)$ also decreases.
    \item From (\ref{end_to_end_delay}), we can see that if $\eta_s$ reduces, $\overline{\eta_{LE}} (with\; \eta_s)$ reduces for a certain LISL range.
    \item For a certain LISL range with a certain $\eta_s$ value, when inter-continental distance increases, $\lambda$ increases. From (\ref{end_to_end_delay}), we can say that $\overline{\eta_{LE}} (with\; \eta_s)$ will increase with the increase of $\lambda$. For longer inter-continental connections, propagation delay as well as node delay (due to more number of hops) is more which increases $\overline{\eta_{LE}} (without\; \eta_s)$ for longer inter-continental connections.
    \item We consider that at LISL range $l_a$ and $l_b$ ($l_b>l_a$), path change rates are $\lambda_a$ and $\lambda_b$, respectively. Also, the average end-to-end latencies without $\eta_s$ are $\overline{\eta_{LE_a}} (without\; \eta_s)$ and $\overline{\eta_{LE_b}} (without\; \eta_s)$, respectively. From Figs. \ref{pathchange} and \ref{latency} values we observe that $\frac{\lambda_a}{\lambda_b}>\frac{\overline{\eta_{LE_a}} (without\; \eta_s)}{\overline{\eta_{LE_b}} (without\; \eta_s)}$. For example, considering New York to Istanbul inter-continental connection, assuming $l_a$=1500 km and $l_b$=1700 km, $\frac{\lambda_a}{\lambda_b}=\frac{37.5}{12.3}=3.049$ and $\frac{\overline{\eta_{LE_a}} (without\; \eta_s)}{\overline{\eta_{LE_b}} (without\; \eta_s)}=\frac{37.9}{36.9}=1.027$. Now, we can write the following:
    \begin{equation}\label{eq7}
        \frac{\lambda_a}{\lambda_b}>\frac{\overline{\eta_{LE_a}} (without\; \eta_s)}{\overline{\eta_{LE_b}} (without\; \eta_s)},
    \end{equation}
    \begin{equation}\label{eq8}
        \frac{\lambda_a\;\eta_s/100}{\overline{\eta_{LE_a}} (without\; \eta_s)}>\frac{\lambda_b\;\eta_s/100}{\overline{\eta_{LE_b}} (without\; \eta_s)},
    \end{equation}
    \begin{equation}\label{eq9}
        \begin{split}
          1+\frac{\lambda_a\;\eta_s/100}{\overline{\eta_{LE_a}} (without\; \eta_s)} & >1+\frac{\lambda_b\;\eta_s/100}{\overline{\eta_{LE_b}} (without\; \eta_s)},\\
          \end{split}
    \end{equation}
    \begin{equation}\label{eq10}
        \begin{split}
        \frac{\overline{\eta_{LE_a}} (without\; \eta_s)+\frac{\lambda_a}{100}\;\eta_s}{\overline{\eta_{LE_a}} (without\; \eta_s)} & >\frac{\overline{\eta_{LE_b}} (without\; \eta_s)+\frac{\lambda_b}{100}\;\eta_s}{\overline{\eta_{LE_b}} (without\; \eta_s)}.\\
        \end{split}
    \end{equation}
Assuming average end-to-end latency with $\eta_s$ as $\overline{\eta_{LE_a}} (with\; \eta_s)$ and $\overline{\eta_{LE_b}} (with\; \eta_s)$ for LISL range $l_a$ and $l_b$, respectively and using (\ref{end_to_end_delay}) we can rewrite (\ref{eq10}) as follows:
\begin{equation}\label{eq11}
    \frac{\overline{\eta_{LE_a}} (with\; \eta_s)}{\overline{\eta_{LE_b}} (with\; \eta_s)} > \frac{\overline{\eta_{LE_a}} (without\; \eta_s)}{\overline{\eta_{LE_b}} (without\; \eta_s)}.
\end{equation}

    % \begin{equation*}
    %     \begin{split}
    %       1+\frac{\lambda_1\;\eta_s}{\overline{\eta_{LE_1}} (without\; \eta_s)} & >1+\frac{\lambda_2\;\eta_s}{\overline{\eta_{LE_2}} (without\; \eta_s)}\\
    %     % \Rightarrow 
    %     \frac{\overline{\eta_{LE_1}} (without\; \eta_s)+\lambda_1\;\eta_s}{\overline{\eta_{LE_1}} (without\; \eta_s)} & > \\
    %     & \frac{\overline{\eta_{LE_2}} (without\; \eta_s)+\lambda_2\;\eta_s}{\overline{\eta_{LE_2}} (without\; \eta_s)}\\
    %     \end{split}
        
    % \end{equation*}
    
    % \begin{equation*}
    %     \begin{split}
    %      \frac{\lambda_1}{\lambda_2} & >\frac{\overline{\eta_{LE_1}} (without\; \eta_s)}{\overline{\eta_{LE_2}} (without\; \eta_s)}\\
    %     \Rightarrow \frac{\lambda_1\;\eta_s}{\overline{\eta_{LE_1}} (without\; \eta_s)} & >\frac{\lambda_2\;\eta_s}{\overline{\eta_{LE_2}} (without\; \eta_s)}\\
    %     \Rightarrow 1+\frac{\lambda_1\;\eta_s}{\overline{\eta_{LE_1}} (without\; \eta_s)} & >1+\frac{\lambda_2\;\eta_s}{\overline{\eta_{LE_2}} (without\; \eta_s)}\\
    %     \Rightarrow
    %     \frac{\overline{\eta_{LE_1}} (without\; \eta_s)}{\overline{\eta_{LE_1}} (with\; \eta_s)} & < \frac{\overline{\eta_{LE_2}} (without\; \eta_s)}{\overline{\eta_{LE_2}} (with\; \eta_s)}\\
    %     \end{split}
    % \end{equation*}
    
\end{itemize}
\subsubsection{Impact of $\eta_s$}
\begin{itemize}
    
    \item We have seen that $\overline{\eta_{LE}} (with\; \eta_s)$ reduces faster compared to $\overline{\eta_{LE}} (without\; \eta_s)$ as LISL range increases. Thus, the ratio $\frac{\overline{\eta_{LE}} (without\; \eta_s)}{\overline{\eta_{LE}} (with\; \eta_s)}$ increases as LISL range increases. From (\ref{impact}), as we can see that $\beta$ is proportional to \big\{$1-\frac{\overline{\eta_{LE}} (without\; \eta_s)}{\overline{\eta_{LE}} (with\; \eta_s)}$\big\}, $\beta$ reduces with the increase of LISL range.
    
    % To show mathematically that $\frac{\overline{\eta_{LE}} (without\; \eta_s)}{\overline{\eta_{LE}} (with\; \eta_s)}$ increases with the increase of LISL range, first we consider that at LISL range $l_1$ and $l_2$ ($l_2>l_1$), path change rates are $\lambda_1$ and $\lambda_2$ respectively. Also, end-to-end latencies without $\eta_s$ are $\overline{\eta_{LE_1}} (without\; \eta_s)$ and $\overline{\eta_{LE_2}} (without\; \eta_s)$ respectively. From Fig. \ref{pathchange} and \ref{latency} values we can say that $\frac{\lambda_1}{\lambda_2}>\frac{\overline{\eta_{LE_1}} (without\; \eta_s)}{\overline{\eta_{LE_2}} (without\; \eta_s)}$. Now, we deduce the following:
    % \begin{equation*}
    %     \begin{split}
    %      \frac{\lambda_1}{\lambda_2} & >\frac{\overline{\eta_{LE_1}} (without\; \eta_s)}{\overline{\eta_{LE_2}} (without\; \eta_s)}\\
    %     \Rightarrow \frac{\lambda_1\;\eta_s}{\overline{\eta_{LE_1}} (without\; \eta_s)} & >\frac{\lambda_2\;\eta_s}{\overline{\eta_{LE_2}} (without\; \eta_s)}\\
    %     \Rightarrow 1+\frac{\lambda_1\;\eta_s}{\overline{\eta_{LE_1}} (without\; \eta_s)} & >1+\frac{\lambda_2\;\eta_s}{\overline{\eta_{LE_2}} (without\; \eta_s)}\\
    %     \Rightarrow
    %     \frac{\overline{\eta_{LE_1}} (without\; \eta_s)}{\overline{\eta_{LE_1}} (with\; \eta_s)} & < \frac{\overline{\eta_{LE_2}} (without\; \eta_s)}{\overline{\eta_{LE_2}} (with\; \eta_s)}\\
    %     \end{split}
    % \end{equation*}
    % where $\overline{\eta_{LE_1}} (with\; \eta_s)$ and $\overline{\eta_{LE_2}} (with\; \eta_s)$ are end-to-end latencies considering $\eta_s$ for LISL range $l_1$ and $l_2$.
    \item $\overline{\eta_{LE}} (with\; \eta_s)$ reduces when $\eta_s$ reduces but $\overline{\eta_{LE}} (without\; \eta_s)$ remains the same which causes the ratio $\frac{\overline{\eta_{LE}} (without\; \eta_s)}{\overline{\eta_{LE}} (with\; \eta_s)}$ to increase. As $\beta$ is proportional to \big\{$1-\frac{\overline{\eta_{LE}} (without\; \eta_s)}{\overline{\eta_{LE}} (with\; \eta_s)}$\big\}, it decreases when $\eta_s$ reduces.
    \item Let us consider that for inter-continental distance $d_x$ and $d_y$ ($d_y>d_x$), path change rates are $\lambda_x$ and $\lambda_y$, respectively. Also, average end-to-end latencies without $\eta_s$ are $\overline{\eta_{LE_x}} (without\; \eta_s)$ and $\overline{\eta_{LE_y}} (without\; \eta_s)$, respectively. From Figs. \ref{pathchange} and \ref{latency} values, we also observe that $\frac{\lambda_x}{\lambda_y}>\frac{\overline{\eta_{LE_x}} (without\; \eta_s)}{\overline{\eta_{LE_y}} (without\; \eta_s)}$ (note that in this discussion, we are varying inter-continental distance, not LISL range). For example, at 1700 km LISL range, for New York to Istanbul and New York to Hanoi inter-continental connection, $\lambda_x$, $\lambda_y$, $\overline{\eta_{LE_x}} (without\; \eta_s)$, and $\overline{\eta_{LE_y}} (without\; \eta_s)$ are 12.3\%, 17.5\%, 36.9 ms, and 64.4 ms, respectively from which we get $\frac{\lambda_x}{\lambda_y}=0.703$ and $\frac{\overline{\eta_{LE_x}} (without\; \eta_s)}{\overline{\eta_{LE_y}} (without\; \eta_s)}=0.573$. Using the approach in (\ref{eq7}) -- (\ref{eq11}), we can come to the conclusion that $ \frac{\overline{\eta_{LE_x}} (without\; \eta_s)}{\overline{\eta_{LE_x}} (with\; \eta_s)} < \frac{\overline{\eta_{LE_y}} (without\; \eta_s)}{\overline{\eta_{LE_y}} (with\; \eta_s)}$  where $\overline{\eta_{LE_x}} (with\; \eta_s)$ and $\overline{\eta_{LE_y}} (with\; \eta_s)$ are average end-to-end latencies considering $\eta_s$ for inter-continental distance $d_x$ and $d_y$, i.e., $ \frac{\overline{\eta_{LE}} (without\; \eta_s)}{\overline{\eta_{LE}} (with\; \eta_s)}$ increases as inter-continental distance increases which reduces $\beta$.
\end{itemize}
\subsubsection{Tolerable Value of $\eta_s$}
\begin{itemize}
    \item (\ref{end_to_end_delay}) represents an equation of a straight line with slope proportional to $\lambda$ considering $\overline{\eta_{LE}} (with\; \eta_s)$ as \emph{y} variable and $\eta_s$ as the \emph{x} variable. As LISL range increases, $\lambda$ decreases which makes the slope of the straight lines to reduce. In addition to $\lambda$, $\overline{\eta_{LE}} (without\; \eta_s)$ also reduces with the increase of LISL range, and from (\ref{tolerableeqn}) we can say that $\eta_{s,\textrm{max}}$ increases with increase in LISL range.\vspace{-0.3em}
\end{itemize}
\subsection{Design Guidelines\vspace{-0.2em}}The values we get from (\ref{tolerableeqn}) are exactly same as we get from intersection points shown in Fig. \ref{tolerablezoom}. Given that $\eta_{LE,\;OFTN}$, $\overline{\eta_{LE}} (without\; \eta_s)$, and $\lambda$ are known for a particular inter-continental connection for a certain LISL range, (\ref{tolerableeqn}) can be used to design LCTs in order to exploit full potential of NNG-FSOSNs. For example:
\begin{itemize}
    \item For New York to London inter-continental connection with 1500 km of LISL range, $\eta_{LE,\;OFTN}$, $\overline{\eta_{LE}} (without\; \eta_s)$, and $\lambda$ are 27.38 ms, 26.6 ms, and 33.9\%, respectively. Putting these values in (\ref{tolerableeqn}), we get $\eta_{s,\textrm{max}}$ as 2.3 ms (same as in Fig. \ref{tolerablezoomlondon}).
    \item With 1700 km LISL range for New York to Istanbul inter-continental connection, putting $\eta_{LE,\;OFTN}$=39.55 ms, $\overline{\eta_{LE}} (without\; \eta_s)$=36.9 ms, and $\lambda$=12.3\% in (\ref{tolerableeqn}), we get $\eta_{s,\textrm{max}}$=21.54 ms, i.e., exactly as shown in Fig. \ref{tolerablezoomist}.
    \item Considering New York to Hanoi inter-continental connection with 5016 km of LISL range, values of $\eta_{LE,\;OFTN}$, $\overline{\eta_{LE}} (without\; \eta_s)$, and $\lambda$ are 64.44 ms, 56.7 ms, and 9.6\%, respectively. Using these values in (\ref{tolerableeqn}), we get $\eta_{s,\textrm{max}}$ equal to 80.63 ms (same as in Fig. \ref{tolerablezoomhanoi}).
\end{itemize}
\section{Conclusion and Future Work}\label{con}
Dynamic LISLs are essential to leverage the full potential of NNG-FSOSNs due to their on-demand flexibility. However, whenever a new LISL is established, LISL setup delay is added to the end-to-end latency. To model the end-to-end latency including LISL setup delay, we study the quantification of LISL setup delay, and calculate the end-to-end latencies for low, medium, and high inter-continental distance connections for different LISL setup delay values. We find that the end-to-end latency depends on path change rate which reduces as LISL range increases but increases as inter-continental distance increases. We also highlight the impact of LISL setup delay on total end-to-end latency which clearly indicates that LISL setup delay cannot be ignored. We observe that the impact of LISL setup delay reduces as LISL range or inter-continental distance increases. We also deduce the formula to find maximum tolerable value of LISL setup delay which represents design guidelines for LCT manufacturers so that FSOSNs can have better latency performance compared to OFTN. We see that for some LISL range, there does not exist any such value of $\eta_{s,\textrm{max}}$. An interesting takeaway point is that higher LISL range has two major benefits. Firstly, highest possible LISL range has the best latency performance. Secondly, it has the highest value of $\eta_{s,\textrm{max}}$ which can be attainable. However, with high LISL range, the penalty is more satellite transmission power and energy consumption. 

It is evident that due to change of shortest path with time slots, LISL setup delay is introduced which negatively impacts the latency of an FSOSN using dynamic LISLs. In order to minimize end-to-end latency, we need to minimize the path change rate so that LISL setup delay is introduced less often. In future, we plan to develop algorithms to minimize the path change rate for a better latency performance.
\section*{Acknowledgment}
This work was supported by the High Throughput and Secure Networks Challenge Program at the National Research Council of Canada. The authors would also like to acknowledge Dr. Pablo Madoery for his technical help and feedback.
\bibliographystyle{IEEEtran}
\footnotesize
% \vspace{-0.5em}
\bibliography{IEEEabrv}
\end{document}